\documentclass{PoS}

\title{Detecting Dark Matter annihilation lines with Fermi}

\ShortTitle{Detecting Dark Matter annihilation lines with Fermi}

\author{\speaker{Tomi Ylinen}\thanks{on behalf of the Fermi-LAT collaboration.} $^{a}$, Yvonne Edmonds$^{b}$, Elliott D. Bloom$^{b}$ and Jan Conrad$^{c}$\\
	\llap{$^{a}$} Department of Physics, Royal Institute of Technology (KTH), AlbaNova, SE-106 91 Stockholm, Sweden \& School of Pure and Applied Natural Sciences, University of Kalmar, SE-391 82 Kalmar, Sweden.\\
	\llap{$^{b}$} Kavli Institute for Particle Astrophysics and Cosmology, SLAC National Accelerator Laboratory, Stanford University, Menlo Park, 94025, USA.\\
	\llap{$^{c}$} Department of Physics, Stockholm University, AlbaNova, SE-106 91 Stockholm, Sweden.\\
        E-mail: \email{tomiy@particle.kth.se}, \email{edmonds@slac.stanford.edu}, \email{bloom@slac.stanford.edu}, \email{conrad@physto.se}}

\abstract{Dark matter constitutes one of the most intriguing but so far unresolved issues in physics today. In many extensions of the Standard Model the existence of a stable Weakly Interacting Massive Particle (WIMP) is predicted. The WIMP is an excellent dark matter particle candidate and one of the most interesting scenarios include an annihilation of two WIMPs into two gamma-rays. If the WIMPs are assumed to be non-relativistic, the resulting photons will both have an energy equal to the mass of the WIMP and manifest themselves as a monochromatic spectral line in the energy spectrum. This type of signal would represent a "smoking gun" for dark matter, since no other known astrophysical process should be able to produce it. In these proceedings we give an overview of the different approaches to a search for dark matter lines that the Fermi-LAT collaboration is pursuing and the various challenges involved.}

\FullConference{Identification of dark matter 2008\\
		 August 18-22, 2008\\
		 Stockholm, Sweden}

\begin{document}

\section{Fermi}
\noindent The Fermi Gamma-ray Space Telescope (``Fermi'')~\cite{FermiWeb}, which was formerly known as GLAST, was launched on June 11, 2008, from Kennedy Space Center in Florida, USA. It is composed of two instruments: the Gamma-ray Burst Monitor (GBM) and  the Large Area Telescope (LAT). LAT is the main instrument for gamma-ray detections with its 4 x 4 array of 16 identical modules, each with a tracker, based on silicon-strip detectors and a calorimeter with CsI(Tl) crystals. The array of trackers is surrounded by a segmented anti-coincidence detector, made of plastic scintillators.

\section{Dark matter lines}
\noindent One of the scientific goals of Fermi is to probe the existence of dark matter. In many extensions of the Standard Model of particle physics, dark matter is predicted to exist in the form of a stable Weakly Interacting Massive Particle (WIMP), denoted here as $\chi$. In many proposed scenarios, WIMPs can annihilate and produce gamma-rays either via hadronization and the subsequent decay of $\pi^{0}$ mesons or through direct annihilation channels to gamma-ray final states. The former gives rise to a continuum of gamma-rays, whereas the latter provides a ``smoking gun'' for dark matter, namely a spectral line from monochromatic gamma-rays. For the $2\gamma$-channel, $E_{\gamma}=m_{\chi}$, and for a $Z\gamma$-channel $E_{\gamma}=m_{\chi}(1 - m^{2}_{\chi}/4\:m^{2}_{Z})$.

In most models, however, the annihilation directly into two gamma-rays is suppressed since this channel is forbidden on tree-level (for a supersymmetric case, see e.g.~\cite{Bergstrom1997}). Thus, the resulting branching fraction is in general quite small, of the order of $10^{-4}$--$10^{-3}$.

There are, however, exceptions where the $2\gamma$-final state is instead the leading channel of annihilations. One example is the Inert Doublet-Model (IDM), which is a minimal extension of the Standard Model with an additional Higgs doublet that has no direct coupling to fermions~\cite{Gustafsson2007}. The inclusion of radiative corrections to dark matter annihilations in the form of internal bremsstrahlung may also produce spectral features similar to a smeared out spectral line~\cite{Bringmann2008}.

\section{Detection strategies}
\noindent The probability to detect a spectral line in the midst of a background depends among other things on the energy resolution of the detector. With simulations, the energy resolution of the LAT has been calculated to be roughly 7\% at 10 GeV and 13\% at 200 GeV~\cite{LATPerformance}. It should be noted also that a monochromatic signal is not a simple Gaussian distribution when seen by the detector but rather a distribution that can be reasonably well fitted with a double-Gaussian function~\cite{Baltz2008}.

One of the primary goals of the Fermi-LAT collaboration in dark matter line searches is therefore to perform an event selection that improves the energy resolution and thereby increases the significance of a potential line signal. The selection must be optimized since a particular event selection may also results in an increased charged particle contamination and a decreased effective area and angular resolution.

A better energy resolution could, however, also be achieved if an alternative and better energy reconstruction algorithm can be developed. This is currently being investigated in the Fermi-LAT collaboration as well.

A second required optimization involves the selection of a region-of-interest (ROI). The ideal ROI depends mainly on two things: the nature of all other sources of gamma-rays, which serve as background to the line signal, and the halo profile of the dark matter, i.e. the distribution of dark matter in the Universe. The reduction of background photons in the signal region can be dealt with by improving the energy resolution as already mentioned, but ROIs based on halo profiles are by definition model-dependent. There are several theoretically motivated ROIs (see e.g.~\cite{Stoehr2003}) but the strategy in the Fermi-LAT collaboration will be to explore multiple scenarios to avoid missing a potential signal. The results presented in this note are based on a standard photon selection, i.e. not optimized for a line search.

\section{Results}
\noindent After the above optimizations, the spectral line search reduces to a statistical problem, to which the background can be determined from the data itself. The significance of a potential line can, however, depend on which statistical method is chosen for the search. Therefore, a variety of statistical methods have been investigated in terms of their power and coverage (see e.g.~\cite{Conrad2007}) and the goal is to have at least two independent line searches to increase the reliability of a potential detection. Examples of the statistical methods that have been studied in the Fermi-LAT collaboration are the $\Delta\chi^{2}$, Scan Statistics~\cite{Terranova2004} and profile likelihood~\cite{Rolke2005}.

\begin{figure}[htb!]
	\centering
		\includegraphics[width=0.7\textwidth]{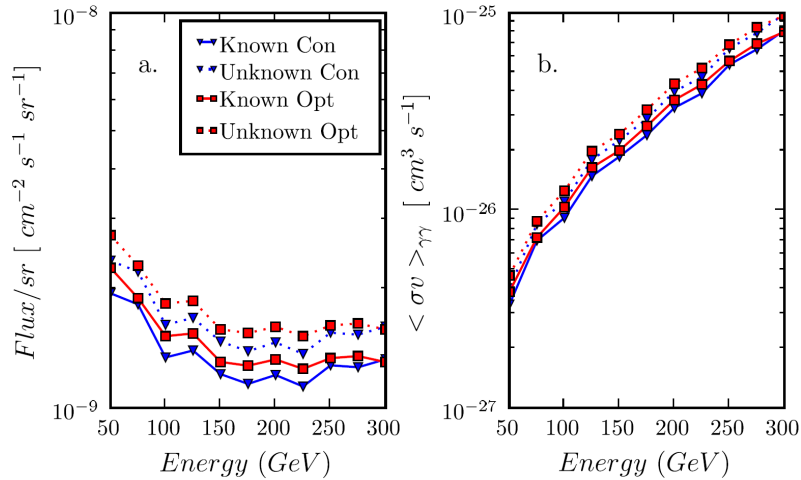}
  \caption[Figure 1]{The LAT $5\sigma$-sensitivity to dark matter line signals in 5 years of operation, determined with simulations and by using the $\Delta\chi^{2}$ method. An NFW halo profile and an ROI given by a broken annulus ($r\in\left[20^{\circ}, 35^{\circ}\right]$, $\left|b\right|>15^{\circ}$) around the Galactic Center are assumed. From~\cite{Baltz2008}.}
	\label{fig1}
\end{figure}

In Fig.~\ref{fig1}, the LAT $5\sigma$-sensitivity to dark matter line signals for 5 years of operation is shown. The limit has been determined using a simulation package developed by the Fermi-LAT collaboration, by applying a $\Delta\chi^{2}$ and by assuming a Navarro-Frenk-White (NFW) halo profile and an ROI given by a broken annulus ($r\in\left[20^{\circ}, 35^{\circ}\right]$, $\left|b\right|>15^{\circ}$) around the Galactic Center~\cite{Baltz2008}. The four lines correspond to two different Galactic diffuse models (conventional~\cite{Strong2000} and optimized~\cite{Strong2004}) in the two cases where the line position is known and unknown.

\begin{figure}[htb!]
	\centering
		\includegraphics[width=0.45\textwidth]{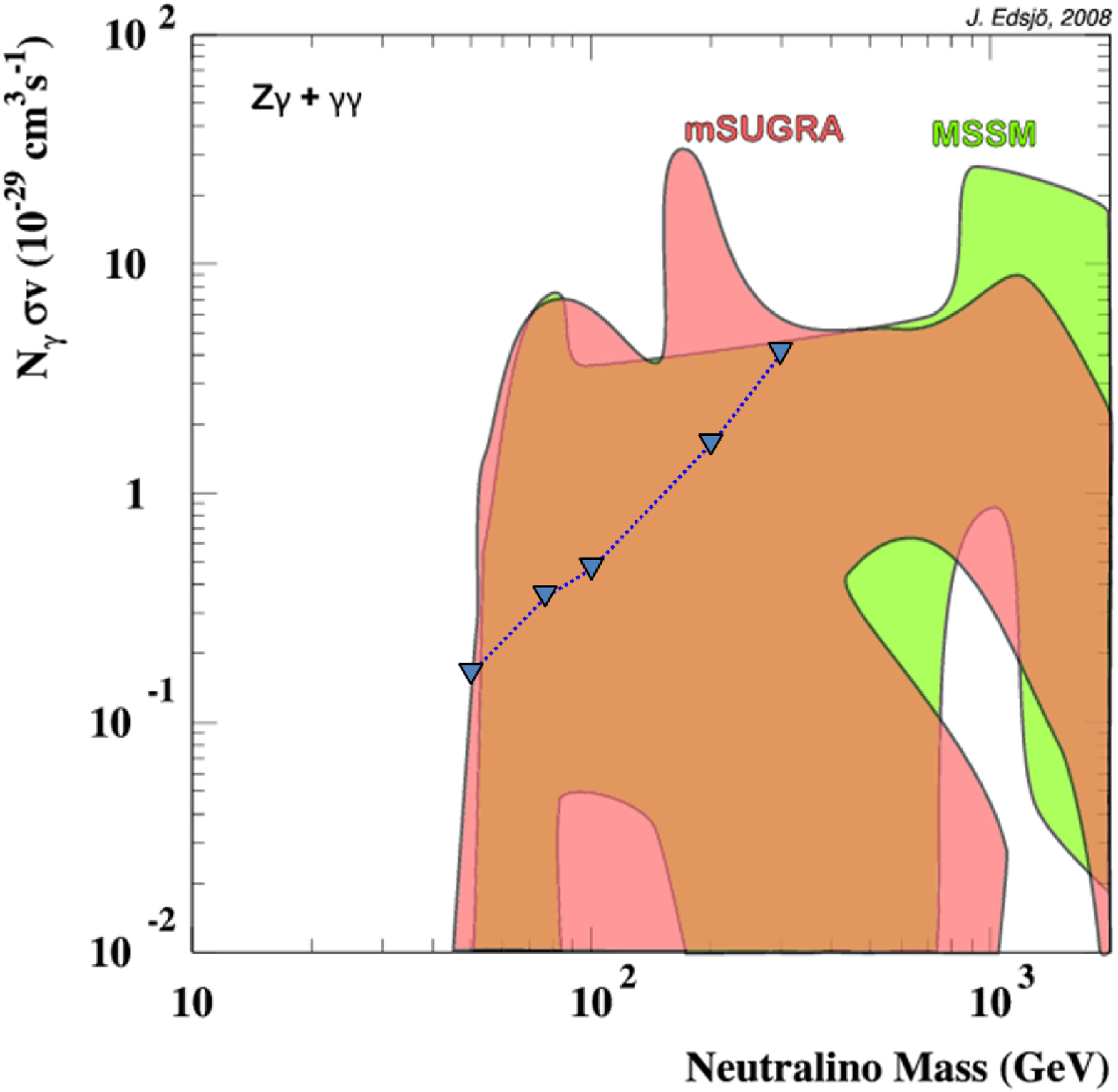}
		\includegraphics[width=0.468\textwidth]{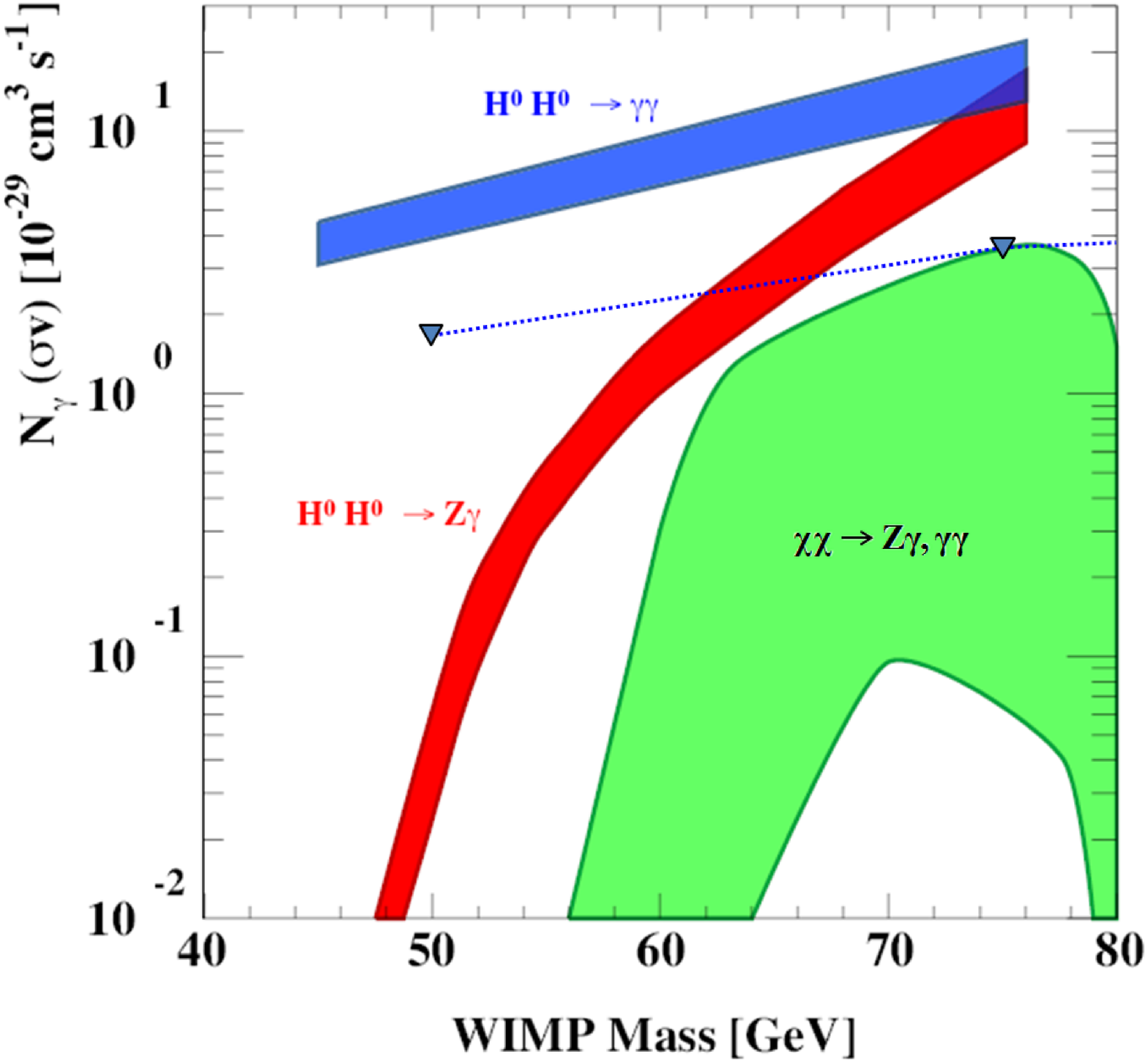}
  \caption[Figure 2]{The best of the upper limits presented in Fig.~\ref{fig1} overlayed on parameter scans for mSUGRA and MSSM, which include accelerator and WMAP constraints (boost factor of $10^{3}$) (\textit{left}), and for IDM (boost factor of $10^{2}$) (\textit{right}).}
	\label{fig2}
\end{figure}

It can be illustrative to compare the best of the upper limits in Fig.~\ref{fig1} to different particle physics models. In the left plot of Fig.~\ref{fig2}, the upper limit has been overlayed on parameter scans for mSUGRA and MSSM that include accelerator and WMAP constraints. The assumed boost factor of $10^{3}$ is admittedly unrealistic for the broken annulus, but might be achieved for other ROIs (e.g. the Galactic Center). The right plot in Fig.~\ref{fig2} shows the upper limit on IDM parameter scans. In this case, a boost factor of $10^{2}$ was assumed.

\end{document}